\documentclass[12pt]{article}
\usepackage{amsmath, amsthm, amssymb,slashed}
\newcommand{\Tr}{{\rm Tr} }
\newcommand{\ev}[1]{\langle{#1}\rangle}

\begin{document}
\begin{titlepage}
\begin{center}
{\bf\Large{\vbox{\centerline{Note on  Mutual Information between Two Intervals of}\vskip .5cm
 \centerline{  Extremal BTZ }}}}
\vskip
0.5cm {Nan Bai$^{a}$ \footnote{bainan@itp.ac.cn}, Yi-Hong Gao$^{a}$ \footnote{gaoyh@itp.ac.cn}, Xiao-bao Xu$^{a}$ \footnote{xbxu@itp.ac.cn}} \vskip 0.05in
{\it ${}^a$ State Key Laboratory of Theoretical Physics,\\
Institute of Theoretical Physics,\\ Chinese Academy of
Sciences, P.O. Box 2735, Beijing 100190, China }

\end{center}
\vskip 0.5in
\baselineskip 16pt
\abstract{In this note we compute mutual information between two intervals in CFTs dual to extremal BTZ (UV CFT) and near horizon limit of extremal BTZ (IR CFT) using the replica technique in some limiting regimes, which can be compared with holographic description. }
\end{titlepage}

\section{Introduction}
Entanglement entropy (EE)  is an important tool in many-body quantum system. Simply speaking, it is the entropy of a subsystem $A$ after tracing out its complement $A^c$. However entanglement entropy  is not easy to compute exactly, only few examples are available \cite{{EE1},{EE2},{EE3}}, where they find that for 2D CFT the entanglement entropy of a single interval $L$ on the plane is
\begin{equation}
S_\mathrm{EE}=\frac{c}{3}\log\frac{L}{\epsilon}~     ,
\end{equation}
where $c$ is the central charge of CFT, and $\epsilon$ is a UV cutoff. In fact we determine it just by using conformal symmetry without knowing the content of CFT. But when we consider the entanglement entropy of a subsystem consists of multiple intervals, the computation becomes difficult because of the well known fact in the Conformal Field Theory that the four point function cann't be determined by only using conformal symmetry, we need to know the full operator content of the theory \cite{{EE4},{EE5},{EE6}}. The difficulty ia also in Renyi entropy, which is the generalization of entanglement entropy, defined as
\begin{equation}
S^{(n)} = \frac{1}{1-n}\ln \Tr \rho^n  \
\end{equation}
where $\rho$ is the reduced density matrix of a subsystem for integer $n\geq2$. Then analytically continue  $n \to 1$  to get the entanglement  entropy
$S_\mathrm{EE}= S^{(1)}=-\Tr \rho \ln \rho \ $. According to replica trick \cite{{EE7},{EE1}}, the Renyi entropy can be calculated by the partition function of the QFT on a $n$-sheeted spacetime which sewing them together cyclically along the entanglement subsystem $A$ at a specific time. In particular, for 2D CFT we may get the partition function through calculating the correlation function of twist operators \cite{{EE2},{EE3},{EE5}}.
We have pointed out the difficulty in calculating the entanglement entropy of multiple intervals, there are still some results in this problem \cite{{EE4},{EE5},{EE6},{EE8},{EE9},{EE10},{EE11}}. In \cite{{EE4}} the author uses the operator product expansion (OPE) of twist operators to compute
the Renyi entropy. Recently, \cite{{EE8},{EE9}} give more properties of the  Renyi entropy for certain 2D dimensional conformal field theories with large central charge by solving the monodromy problem numerically. And \cite{{EE10}} obtain the one loop correction to  the Renyi entropy using the method in \cite{{EE9}}. Remarkably, a more explicit computation of the Renyi entropy in OPE now is attained \cite{{EE11}}, it can give more quantum corrections to
the Renyi entropy than \cite{{EE10}}.

Entanglement entropy, like the Wilson loops \cite{{WL}}, have a holographic description \cite{{EE12}}, Ryu and Takayanagi (RT) proposed that the entanglement entropy of an arbitrary spatial region on the boundary  is given by the area of a minimal surface in the bulk. More specifically,
\begin{equation}
S_A=\mbox{min}_\gamma \frac{\mbox{area}\left(\gamma\right)}{4 G_N}
\end{equation}
where $\gamma$ is the static surface in the bulk whose boundary is $\partial A$ with Newton constant $G_N$.
There are many discussions on the holographic proposal of entanglement entropy \cite{{EE13},{EE14},{EE15},{EE16},{EE17}}. In particular, \cite{{EE13}} extended the proposal to holographic dual of entanglement entropy in a time varying QFT, called covariant holographic entanglement entropy. Their proposal
is that the entanglement entropy in time-dependent background is given by the smallest one of the area of \emph{extremal surface}, this has been applied
to holographic thermalization \cite{{EE18}}.

Very recently, \cite{{EE19}} computed the covariant holographic entanglement entropy of a single interval in the context of warped AdS$_3$ background, which is the near horizon geometry of the extremal BTZ black hole, their result is in agreement with the one of the rotating BTZ black hole \cite{{EE13}}. And \cite{{EE20}} studied the entanglement entropy of a single interval in the CFT at the boundary of extremal BTZ and at the boundary of the near-horizon limit of extremal BTZ, they find both  the entanglement entropy can be written  in one formula. In this note we attempt to study the entanglement entropy of two intervals for both CFT in \cite{{EE20}} and investigate the mutual information. We want to see whether the result in \cite{{EE20}} still holds.
However we find it is hard to compute the four point function of twist operators even using OPE expansion like in \cite{{EE4},{EE8},{EE11}}. We calculate the mutual information of two intervals in the specific limit  and argue that they are consistent with the gravity duals. It is desirable to calculate the the mutual information of two intervals more generally without in the specific situation, then we can see how the entanglement entropy changed in the near horizon limit of extremal BTZ black hole, which may help us understanding better the mechanisms of entanglement entropy under holographic renormalization \cite{{EE21},{EE22}}.

To make the note more readable, in section 2 we briefly review the results in \cite{{EE20}}. Then we get the mutual information of two intervals for both CFTs under specific limit in the next section. The last section is some discussion.

\section{Entanglement entropy of CFTs dual to extremal BTZ}
We recall the known fact about BTZ black hole which is already presented in \cite{{EE20}}. The metric of extremal BTZ black hole is
\begin{equation}
\mathrm{d}s^2=-\frac{(r^2-r^2_0)^2}{R^2r^2}dt^2+\frac{R^2r^2}{(r^2-r^2_0)^2}dr^2+r^2(d\phi-\frac{r^2_0}{Rr^2}dt)^2
\label{extremal-metric}
\end{equation}
where $R$ is the radius of anti-de Sitter space and $r_0$ is the horizon of extremal BTZ. It is easy to see that the Hawking temperature vanishes.  And the metric  can be mapped to Poincare patch of AdS$_3$
\begin{equation}
\mathrm{d}s^2=R^2\frac{dU^2+d\omega_+ d\omega_-}{U^2}
\label{AdS3}
\end{equation}
by the coordinate transformation
\begin{equation}
w_- = \frac{R}{2r_0}e^{\frac{2r_0}{R}(\phi-t/R)} ~, \qquad
w_+ = \phi+\frac{t}{R}-\frac{Rr_0}{r^2-r^2_0} ~, \qquad
U = \frac{R}{\sqrt{r^2-r^2_0}}e^{\frac{r_0}{R}(\phi-t/R)} \label{mapE}
\end{equation}
We call the CFT on the boundary of the extremal BTZ black hole \eqref{extremal-metric} \emph{ultraviolet} conformal field theories (UV CFT) according to AdS$_3$/CFT$_2$.

The near horizon geometry of extremal BTZ \footnote{\cite{{EE20}} give the explicit derivation of the geometry} is written in a warped AdS$_3$ form as
\begin{equation}
\mathrm{d}s^2=\frac{R^2}{4}\,\left(-\rho^2\,dv_+^2+\frac{d\rho^2}{\rho^2}\right)+{R^2}\,\left(dv_- +\frac{\rho}{2}\, dv_+\right)^2  , \qquad
v_-\sim v_-+2\pi \frac{r_0}{R} \label{NH}
\end{equation}
It should be point out that  $\rho$ and $v_\pm$ are dimensionless. The near horizon geometry of extremal BTZ also has a CFT dual, we refer to it as  infrared CFT (IR CFT). The metric can also be bringed to Poincare patch of AdS$_3$ through the coordinate transformation
\begin{equation}
\omega_-=\frac1{2}\,e^{2v_-} , \qquad
\omega_+=v_+-\frac{1}{\rho} , \qquad
U=\frac{e^{v_-}}{\sqrt{\rho}} \label{mapNH}
\end{equation}
It have been noted that the transformation \eqref{mapE} and \eqref{mapNH} are similar to each other in \cite{EE21}, this lead to the entanglement entropy
within the ultraviolet CFT matches the entanglement entropy within the infrared CFT, which is proved below.

We can now compute the entanglement entropy of a single interval, \emph{i.e.} the two-point function of the twist operators, however because both CFTs live on the cylinder, we need the following formula to calculate the correlations on the cylinder
\begin{equation}
\langle {\cal O}(z_1,\bar{z}_1) {\cal O}(z_2,\bar{z}_2)\ldots \rangle=
\prod_i \left(\frac{dw_i}{dz_i}\right)^h \left(\frac{d\bar w_i}{d\bar z_i}\right)^{\bar h} \langle  {\cal O}(w_1,\bar{w}_1) {\cal O}(w_2,\bar{w}_2) \ldots \rangle
\label{corr-trans}
\end{equation}
where $w(z)$ is the map from cylinder to plane.

Making  $r\to \infty$ and Wick rotating $t\to i\tau$, we get the map
\begin{eqnarray}
w(z)&\equiv&\phi+i\frac{\tau}{R}= z ~, \nonumber \\
\bar{w}(\bar{z})&\equiv& \frac{R}{2r_0}e^{\frac{2r_0}{R}(\phi-i\tau/R)}= \frac{R}{2r_0}e^{\frac{2r_0}{R}\bar{z}}\label{Trans}
\end{eqnarray}
We then derive the two-point function of the twist fields in the ultraviolet CFT \footnote{The derivation is given in the appendix A of \cite{EE20}}
\begin{equation}
\langle \phi^+(z_1,\bar{z}_1)\phi^-(z_2,\bar{z}_2)\rangle
=\left(\frac{z_2-z_1}{\epsilon}\right)^{-2h_n}\left(\frac{R}{r_0\epsilon}\sinh\left(\frac{r_0}{R}(\bar{z}_2-\bar{z}_1)\right)\right)^{-2\bar{h}_n}
\end{equation}
where the conformal dimension of the twist fields  $h_n,\bar{h}_n$ is equal to
\begin{equation}
h_n=\bar{h}_n=\frac{c}{24}\left(n-\frac1{n}\right)
\end{equation}
So the entanglement entropy of a single length $L$ interval is
\begin{equation}
S_\mathrm{EE}=\frac{c}{6}\log\left(\frac{L}{\epsilon}\right)+\frac{\bar{c}}{6}\log\left(\frac{R}{r_0\epsilon}\sinh\left(\frac{r_0\,L}{R}\right)\right) \label{eBTZEE}
\end{equation}

We can see directly  the analogy between \eqref{mapNH} and \eqref{Trans}, therefore we get the entanglement entropy of a length $L$ interval within the IR CFT quickly
\begin{equation}
S_\mathrm{EE}=\frac{c}{6}\log\left(\frac{L}{\epsilon}\right)+\frac{\bar{c}}{6}\log\left(\frac{1}{\epsilon}\sinh\left(L\right)\right)  \label{NHZEE}
\end{equation}

We can check the RT conjecture (covariant holographic entanglement entropy proposal) by this simple instance.  First the formula \eqref{eBTZEE} is inentical to the formula (2.22) of \cite{EE19} which is obtained by taking the extremal limit of covariant holographic  entanglement entropy of rotating BTZ black hole \cite{EE13} with the equality $\beta_R=\frac{\pi R}{r_0}$. Second we can reduce the formula (2.9) of covariant holographic  entanglement entropy  for warped AdS$_3$ \cite{EE19}  to the formula \eqref{NHZEE} if we take $l=R$, $r_+=r_0$, $\frac{\phi}{2 l}=v_-$, $\psi=v_+$. \footnote{We don't show that the formula \eqref{NHZEE} is equal to the formula (2.9) of \cite{EE19} exactly, we think that the difference is caused by the choice of the UV cut off $\epsilon$ in the two calculation, and we don't worry that it is a serious problem to checking the RT conjecture.}

\section{Mutual Information for two disjoint intervals}
In this section, we will compute mutual information of both CFTs dual to extremal BTZ and near horizon limit of extremal BTZ by employing the above result in the some limiting regimes.  We consider two disjoint intervals A and B, the mutual information between disjoint regions A and B is defined as $I_{A:B}\equiv S_A+S_B-S_{A\bigcup B}$.

There are some properties of the mutual information
\begin{enumerate}
\item  $ I_{A:B}\geq 0$, due to the entanglement entropy satisfies strong subadditivity, i.e.
\begin{equation}
S_A+S_B\geq S_{A\bigcup B}
\end{equation}
\item  Mutual Information measure the amount of correlation, both classical and quantum, between two subsystems A and B, and  set an upper  bound on quantum correlators  \cite{{EE23}}
\begin{equation}
I_{A:B}\geq  \frac{\left(\ev{\mathcal{O}_A}\ev{\mathcal{O}_B}  -\ev{\mathcal{O}_A \mathcal{O}_B}\right)^2}{2 || \mathcal{O}_A ||^2 || \mathcal{O}_B ||^2 }
\end{equation}
for an operator $\mathcal{O}_A$ in the region A and  $\mathcal{O}_B$ in the region B.
\item Unlike entanglement entropy, mutual information is free of ultraviolet divergence in 2D CFT \cite{{EE2}}, it is believed to be right in higher dimensions.
\end{enumerate}
Using the RT proposal, Headrick  find a first-order phase transition for holographic mutual information, it has a discontinuous first derivative as a function of the separation between A and B \cite{{EE4}}, more explicitly
\begin{equation}
I_{A:B}(x) = \begin{cases} 0\,,\quad & x\le1/2 \\ (c/3)\ln(x/(1-x))\,,\quad & x\ge1/2\end{cases}
\end{equation}
where x is the conformal four point ratio defined by the size and the separation of intervals A and B.
In \cite{{EE4}} Headrick argued that, the first order phase transition of mutual information occurs since the global minimum changes between the local minima, as we vary $x$, and is analogous to the Hawking-Page transition which is due to competing saddle points of the Euclidean action.

Now we want to calculate the mutual information of two intervals A and B in both CFTs dual to extremal BTZ and near horizon limit of extremal BTZ, mainly we need to calculate the entanglement entropy of two intervals A and B, i.e. the four point correlation  function of the twist operators. However as we already stated in the introduction, it is difficult to compute  the four point correlation in general CFT, which is also a  known problem in  Conformal Field Theory, see discussions in  \cite{{cft1}}.

We consider two identical intervals A and B of length $L$ separated by a distance $l$ in the ultraviolet CFT dual to extremal BTZ,  the reduced density matrix is given by a 4-point function of twist fields,
\begin{eqnarray}
\Tr{\rho^n} =\ev{\Phi^+(z_1,\bar{z}_1)\Phi^-(z_2,\bar{z}_2)\Phi^+(z_3,\bar{z}_3) \Phi^-(z_4,\bar{z}_4)}  \label{fourpoint}\\
\quad z_1 = \bar{z}_1 = 0 , \quad z_2 =\bar{z}_2 = L ,\quad z_3 = \bar{z}_3=L+l , \quad z_4 =\bar{z}_4 =2L+l \label{poisition}
\end{eqnarray}
In general we can take arbitrary values for $z_i$, $\bar{z}_i$.

Once given the the reduced density matrix $\Tr{\rho^n}$, the entanglement entropy is
\begin{equation}
S_\mathrm{EE}=-\lim_{n\to 1}\partial_n \Tr(\rho^n)
\end{equation}

Now we can use the formula \eqref{corr-trans} and the coordinate transformation \eqref{mapE} mapping 4-point function \eqref{fourpoint} on the cylinder of the boundary of extremal BTZ   to the $w$ plane
\begin{eqnarray}
\ev{\Phi^+(z_1,\bar{z}_1)\Phi^-(z_2,\bar{z}_2)\Phi^+(z_3,\bar{z}_3) \Phi^-(z_4,\bar{z}_4)}\nonumber \\=e^{\frac{2 r_0}{R}\left(2 L+l\right)\Delta_n}
\ev{\Phi^+(w_1,\bar{w}_1)\Phi^-(w_2,\bar{w}_2)\Phi^+(w_3,\bar{w}_3) \Phi^-(w_4,\bar{w}_4)} \label{4foint1}
\end{eqnarray}
where $\Delta_n=\bar{\Delta}_n\equiv2 h_n$.

Furthermore we can rewrite the  4-point function of twist fields on the $w$  plane according to global conformal symmetry \cite{{cft2}} as
\begin{equation}
\ev{\Phi^+(w_1,\bar{w}_1)\Phi^-(w_2,\bar{w}_2)\Phi^+(w_3,\bar{w}_3) \Phi^-(w_4,\bar{w}_4)}= \left(w_{1 2}w_{3 4}\bar{w}_{1 2}\bar{w}_{3 4}\right)\left(x \bar{x}\right)^{\Delta_n} G_n\left(x,\bar{x}\right) \label{4foint2}
\end{equation}
where $x$ is the conformal four-point ratio defined as
\begin{equation}
x=\frac{\left(w_1-w_2\right)\left(w_3-w_4\right)}{\left(w_1-w_3\right)\left(w_2-w_4\right)} \label{crossratio}
\end{equation}
and $w_{1 2}=w_1-w_2, \bar{w}_{3 4}=\bar{w}_3-\bar{w}_4$ etc.
$G_n\left(x,\bar{x}\right)$ is a function of the cross ratio $x$, and it is invariant under the global conformation transformation.
However we can't compute the  $G_n\left(x,\bar{x}\right)$ analytically always  due to it depends on the full operators content of conformal field theory, which can be trackable in expansion of $x$ through so called "conformal block", see more details in \cite{{cft2}}.

So from \eqref{crossratio} we have
\begin{eqnarray}
x&=&\frac{L^2}{\left(L+l\right)^2}\\
\bar{x}&=&e^{\frac{2r_0}{R} l}\frac{\left(e^{\frac{2r_0}{R}L}-1\right)^2}{\left(e^{\frac{2r_0}{R}\left(L+l\right)}-1\right)^2}
\end{eqnarray}
Combine \eqref{4foint1} and \eqref{4foint2} we get
\begin{equation}
\ev{\Phi^+(z_1,\bar{z}_1)\Phi^-(z_2,\bar{z}_2)\Phi^+(z_3,\bar{z}_3) \Phi^-(z_4,\bar{z}_4)}=\left(\frac{L R}{r_0} \sinh\frac{r_0 L}{R}\right)^{-2 \Delta_n} \left(x \bar{x}\right)^{\Delta_n} G_n\left(x,\bar{x}\right) \label{new4point1}
\end{equation}

Because the similarity between transformation \eqref{mapE} and \eqref{mapNH} , we deduce the reduced density matrix of the CFT dual to near horizon limit of extremal BTZ, we still choose the positions of the two intervals as \eqref{poisition}
\begin{equation}
\ev{\Phi^+(z_1,\bar{z}_1)\Phi^-(z_2,\bar{z}_2)\Phi^+(z_3,\bar{z}_3)\Phi^-(z_4,\bar{z}_4)}=\left(L \sinh L\right)^{-2 \Delta_n} \left(x \bar{x}\right)^{\Delta_n} K_n\left(x,\bar{x}\right) \label{new4point2}
\end{equation}
where $x$ is not changed, $\bar{x}$ now is changed to
\begin{equation}
\bar{x}=e^{2 l}\frac{\left(e^{2 L}-1\right)^2}{\left(e^{2\left(L+l\right)}-1\right)^2}
\end{equation}
We may conjecture more explicitly that the function $G_n\left(x,\bar{x}\right)$ in \eqref{new4point1} and $K_n\left(x,\bar{x}\right)$ in \eqref{new4point2} should have the same structure in the variables $x$, $\bar{x}$.

The conjecture in the above should be right if the ultraviolet CFT dual to extremal BTZ and the infrared CFT dual to near horizon limit of extremal BTZ
belong to the same class of conformal field theory intuitively, but we don't know how the near horizon limit deform the ultraviolet CFT.  As  has been pointed out in \cite{{EE20}}, the similarity between the entanglement entropy in both CFTs can be followed from the  analogy between the transformation \eqref{mapNH} and \eqref{Trans}. So it is desirable to study the relation between the ultraviolet CFT dual to extremal BTZ and the infrared CFT dual to near horizon limit of extremal BTZ more deeply to give a definite conclusion of the entanglement entropy of both CFTs.

Although it is hard to compute the 4-point function, we can still calculate it easily in some limiting regime.  See also discussions in \cite{{EE24}}.
For example we compute the 4-point function \eqref{fourpoint}. When $x\rightarrow 0$, the 4-point function factorizes to \cite{{cft3}},
\begin{eqnarray}
\ev{\Phi^+(z_1,\bar{z}_1)\Phi^-(z_2,\bar{z}_2)\Phi^+(z_3,\bar{z}_3)\Phi^-(z_4,\bar{z}_4)}=\nonumber\\
\ev{\Phi^+(z_1,\bar{z}_1)\Phi^-(z_2,\bar{z}_2)}\ev{\Phi^+(z_3,\bar{z}_3)\Phi^-(z_4,\bar{z}_4)} \label{facr1}
\end{eqnarray}
$x\rightarrow 1$, the 4-point function factorizes to,
\begin{eqnarray}
\ev{\Phi^+(z_1,\bar{z}_1)\Phi^-(z_2,\bar{z}_2)\Phi^+(z_3,\bar{z}_3)\Phi^-(z_4,\bar{z}_4)}=\nonumber\\
\ev{\Phi^+(z_1,\bar{z}_1)\Phi^-(z_4,\bar{z}_4)}\ev{\Phi^+(z_2,\bar{z}_2)\Phi^-(z_3,\bar{z}_3)}  \label{facr2}
\end{eqnarray}
From the formula \eqref{facr1}, we immediately get that the mutual information between two intervals when $x\rightarrow 0$  is zero, which correspond to the decouple of two intervals.

From the formula \eqref{facr2}, we have the mutual information between two intervals when $x\rightarrow 1$ as
\begin{equation}
I_{A:B}=\frac{c}{6}\ln\frac{L^2}{l\left(2 L+l\right)}+\frac{c}{6}\ln\frac{\left(\sinh\frac{r_0 L}{R}\right)^2}{\sinh\frac{r_0 \left(2 L+l\right)}{R}\sinh\frac{r_0 l}{R}} \label{re1}
\end{equation}
This should also be  the result of holographic mutual information using the RT proposal, like  \cite{{EE4}} have done.

Then the mutual information  in the  CFT dual to near horizon limit of extremal BTZ is
\begin{equation}
I_{A:B}(x) = \begin{cases} 0\,,\quad & x\to 0 \\ \frac{c}{6}\ln\frac{L^2}{l\left(2 L+l\right)}+\frac{c}{6}\ln\frac{\left(\sinh L\right)^2}{\sinh \left(2 L+l\right)\sinh l}\,,\quad & x\to 1\end{cases} \label{re2}
\end{equation}

\section{Conclusion}
In this short note we compute the mutual information of two intervals in both CFTs dual to extremal BTZ and near horizon limit of extremal BTZ by using  conformal symmetry. We give the result of the mutual information in some limiting regimes. Our results \eqref{re1} and \eqref{re2} should be the same as the holographic ones because we have checked the entanglement entropy of a single interval between CFT side and holographic dual side. We can't say more about the similarity of the entanglement entropy of two intervals between both CFTs than observation  in \cite{{EE20}}. We need to study the RG flow
from ultraviolet CFT to infrared CFT \cite{{rg}} how  change  $G_n\left(x,\bar{x}\right)$.  This may help us understand the relation between entanglement entropy and renormalization \cite{{EE21},{EE22}}.

\end{document}